# Development of an EMT model of the Balearic power system

**Yousef PIPELZADEH**
**MHI**
**UK**
**ypipelzadeh@mhi.ca**

**Dharshana MUTHUMUNI**
**MHI**
**Canada**
**dharsshana@mhi.ca**

**Farid MOSALLAT**
**Canada**
**faridmos@gmail.com**

**Javier RENEDO***
**Red Eléctrica – Redeia**
**Spain**
**javier.renedo@ree.es**

**Silvia SANZ VERDUGO**
**Red Eléctrica – Redeia**
**Spain**
**sisanz@ree.es**

**Antonio CORDÓN**
**Red Eléctrica – Redeia**
**Spain**
**acordon@ree.es**

**Edgar NUÑO**
**Red Eléctrica – Redeia**
**Spain**
**edgar.nuno@ree.es**

**Macarena MARTÍN**
**Red Eléctrica – Redeia**
**Spain**
**macarena.martin@ree.es**

**SUMMARY**
The Balearic Islands are striving to achieve 100% renewable energy, which poses new challenges in operating the power system securely and reliably. With a higher integration of inverter-based resources (IBRs) and a reduced presence of conventional synchronous generators, the system strength, particularly short circuit levels, becomes weaker and more susceptible to disturbances; and the total inertia of the system becomes lower. Technology enablers are planned to achieve energy transition in the Balearic power system: a new 2x200 MW VSC-HVDC link (bipole with metallic return), Synchronous Compensators (SC) and Battery Energy Storage Systems (BESS), as fully-integrated network components. Detailed ElectroMagnetic Transient (EMT) simulation studies may be needed to analyse stability of the system, due to the high amounts of power-electronics devices in the system. This paper describes the implementation of an EMT model of the Balearic power system.

**KEYWORDS**
Electromagnetic transient (EMT), Root-Mean-Square (RMS) simulation, power system stability, HVDC, voltage source converter (VSC), inverter-based resources (IBR), High Performance Computing (HPC), renewable energy, grid-forming.

This is an unabridged preprint of the following paper:

- Y. Pipelzadeh, D. Muthumuni, F. Mosallat, J. Renedo, S. Sanz Verdugo, A. Cordón, E. Nuño, M. Martín, "Development of an EMT model of the Balearic power system," *Proc. CIGRE Paris Session, B4-PS3, paper 10624*, Paris, France, 25-30 Aug. 2024, pp. 1-13.

## 1. INTRODUCTION

The Spanish Transmission System Operator (TSO) plans to build a new High Voltage Direct Current (HVDC) interconnection between the Iberian Peninsula power system and the Balearic Islands power system. The HVDC link will be based on Voltage Source Converter (VSC) technology and will be a bipole of 2x200 MW and ±250 kVdc of DC voltage. One converter station will be connected to El Fadrell 400 kV substation (Castellón, Valencia, Iberian Peninsula), while the other converter station will be connected to San Martín 220 kV substation (Mallorca Island, Balearic Islands). The link will have HVDC submarine cables of 363km (approx.). The new VSC-HVDC link will have grid-forming control in Mallorca substation, to be able to operate a decarbonized Balearic power system, maintaining the reliability and security of the system.

The Balearic Islands, including Mallorca, Menorca, Ibiza, and Formentera, are a Spanish Archipelago situated in the Mediterranean Sea off the eastern coast of the Iberian Peninsula, as shown in Figure 1.
The four islands are interconnected through various AC submarine cables, forming one unique synchronous electrical system. The peak demand of Balearic power system expected by 2026 is around 1271 MW, while the peak demand of the Iberian Peninsula power system, estimated for 2026, is around 46800 MW. Currently, the Iberian Peninsula and the Balearic Islands are interconnected through a 2x200 MW ±250 kVdc HVDC based Line Commutated Converters (LCC-HVDC) link.

The decarbonization of Balearic Islands will be driven by two main aspects: (a) integration of solar PV into the Balearic power system and (b) Renewable Energy Source (RES) production of the Iberian Peninsula transferred to Balearic Islands through the HVDC interconnections. This will imply operating the system without synchronous generators in certain scenarios. To enable high penetration of renewable energy sources and ensure system stability, several measures have been planned, including the deployment of the new VSC-HVDC link, synchronous compensators (SC) and battery energy storage systems (BESS) as fully-integrated network components, according to Article 63 of European Union (UE) Directive 2019/944 [2]. All details of the planning of the Spanish power system in horizon H2026 can be found in [1].

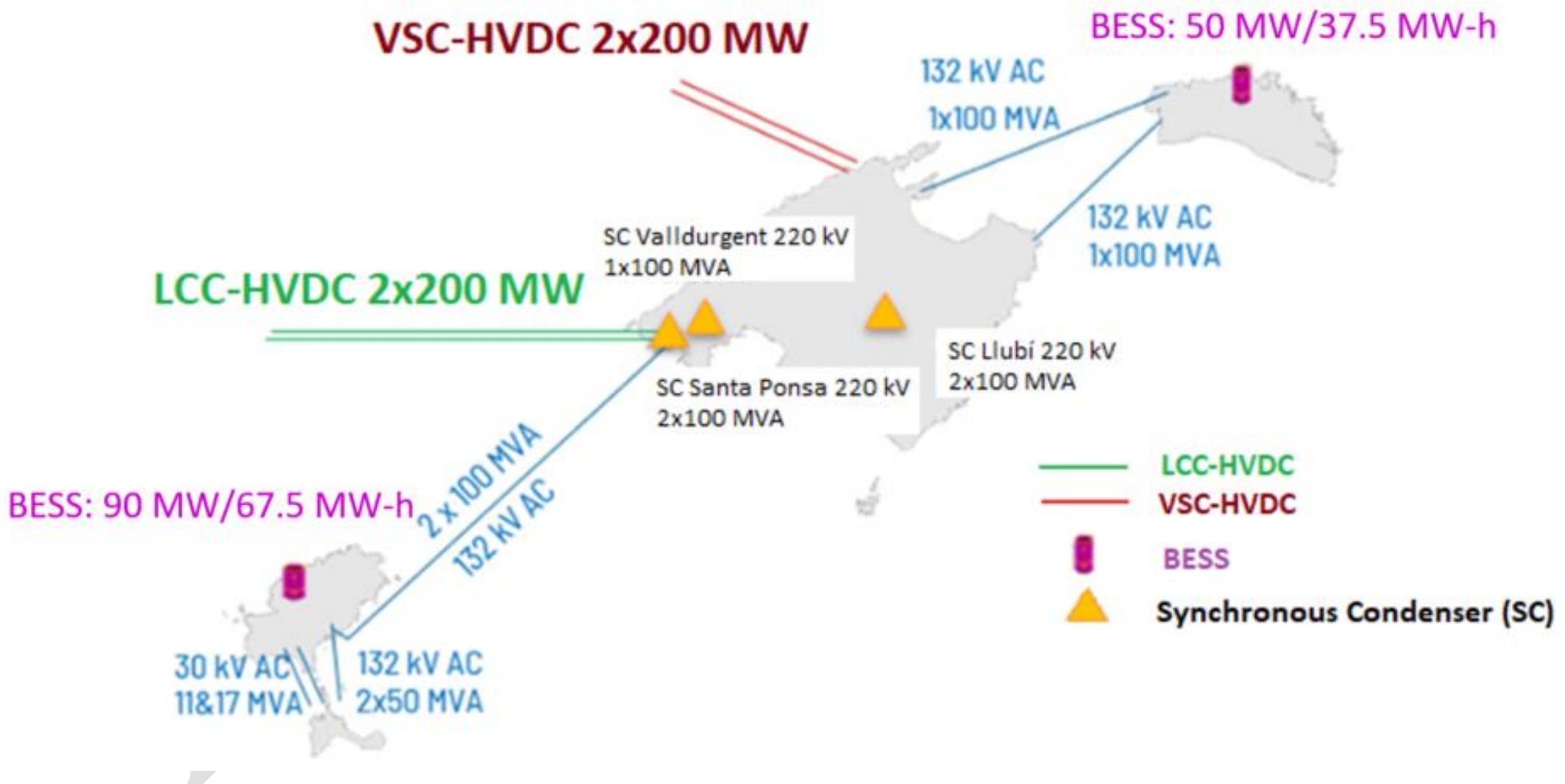

Figure 1: The Balearic Islands, including Mallorca, Menorca, Ibiza, and Formentera. H2026.

The new VSC-HVDC interconnection project presents several challenges, including operating the link in different scenarios, coexistence with the existing LCC-HVDC link, defining robust control and protection schemes, and ensuring no interactions with nearby devices. To address these challenges, a comprehensive Electromagnetic Transient (EMT) simulation model representing the entire Balearic Islands enabling large-scale system level EMT studies is developed and presented in this paper. This project delivers an EMT model of the Balearic Islands for H2026 and H2030 including the 2 HVDC links, PV plants, BESS and Synchronous Compensators.

The adoption of EMT simulation in large-scale power systems is driven by the evolving complexity of modern electrical networks, particularly with the increased integration of IBRs and advanced HVDC technologies. The need for detailed EMT stability studies in large-scale power systems with large amounts of inverter-based resources [3].
This shift is not just theoretical but is being practically implemented by several TSOs globally, including those in the UK [4], Australia [5] and several European operators, among others. These operators are developing large-scale EMT models of their networks, leveraging advancements in fast computing and model parallelization techniques that yield efficient run times. EMT's ability to deliver detailed, high-fidelity modeling is crucial for capturing the transient behaviors of modern grid components, ensuring system stability and the effective integration of renewable energy. This approach enables a deeper understanding of system dynamics, essential for informed decision-making and strategic planning. Additional examples of EMT studies in large-scale power systems can be found in [6], [7], [8], [9].

This paper report focuses on the development of a comprehensive model for the Balearic Islands power system suitable for EMT simulation in a commercial software [10].

The Spanish TSO is dedicated to developing a comprehensive EMT environment for operational and planning simulations. This paper addresses critical components and methodologies essential for achieving this objective, which include:

- **Methodology for Transitioning the Balearic Island Power System to an EMT Simulation Framework**: This section outlines the systematic approach for converting the Balearic Island power system from a conventional RMS simulation model to an EMT simulation framework.

- **Development of a large-scale EMT Model for the Balearic Islands**: A comprehensive EMT model, which incorporates large-scale inverter-based resources, reflecting the complex dynamics of the Balearic Islands power system.

- **VSC-HVDC Link for Interconnection Between the Balearic Islands and the Iberian Peninsula**: The paper discusses the development and specification of the VSC-HVDC link that connects the Balearic Islands to the mainland, highlighting its operational performance within the EMT environment. The VSC-HVDC link model will have the option of grid-following (GFL) or grid-forming (GFM) control modes at Mallorca converter station.

- **Network Splitting Strategies in EMT for Enhanced Parallel Simulation Performance**: Techniques for dividing the network within the EMT simulations are highlighted and drawing attention to parallelization of complex wide-area models can significantly improve simulation speed and efficiency.

- **Steady-State and Dynamic Validation Across RMS and EMT Environments**: Dynamic validation between the original Root-Mean-Square (RMS) models and the developed EMT models, emphasizing the accuracy of system behavior.

In the past, a reduced model of the Balearic power system was also developed within a Research & Development (R&D) project as a preliminary study to analyse the impact of control schemes of the new VSC-HVDC system on power system stability [11]. The differences with the model presented in this work are that the model of the Balearic power system contains significant more details. The entire process of the conversion from RMS data to EMT data is covered and a detailed validation of the proposed model is presented.

## 2. METHODOLOGY FOR TRANSITIONING THE BALEARIC ISLAND POWER SYSTEM TO AN EMT SIMULATION FRAMEWORK

This section details the conversion of the Balearic Island power system from an RMS environment to an EMT modelling framework. The process encompasses a flow of steps, as depicted in Figure 2, which collectively ensure a comprehensive and accurate transition from the initial system representation in RSM environment to an EMT environment.

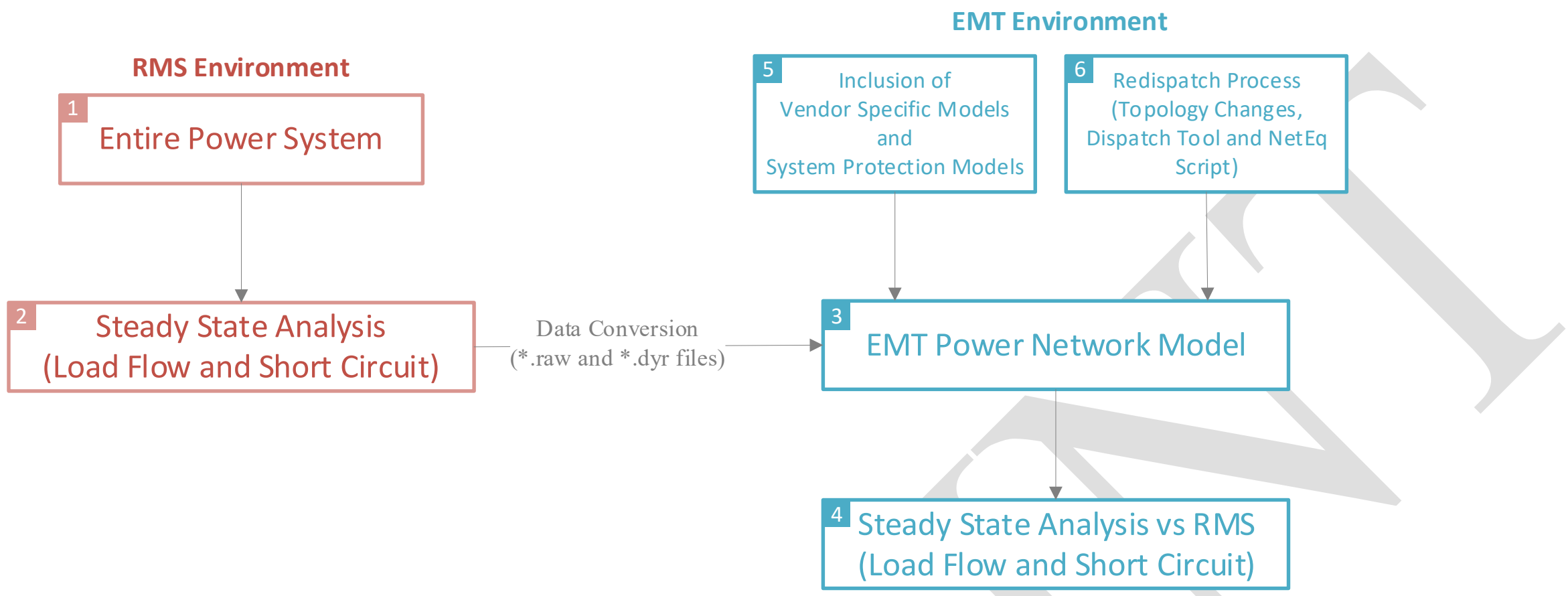

Figure 2: Workflow Diagram for the Conversion of the Balearic Island Power System from RMS to EMT Environment

**Entire Power System (RMS Environment):** The Balearic Island network model, managed by the Spanish TSO, is within RMS environment. This model includes the full electrical network (as well as dynamic models).

**Steady State Analysis (Load Flow):** A steady-state analysis is conducted in the RMS environment to evaluate the load flow. The load flow solution forms the baseline for initializing and operating condition, essential for subsequent data conversion.

**Data Conversion:** the data conversion step employs tools such as a power system importer (PRSIM) and uses the RMS data files (PSS/E data from '*.raw' and '.dyr' files) as the data input and facilitates the development of the EMT environment. This critical procedure ensures that all network details, are accurately mapped onto the EMT model.

**EMT Power Network Model:** The Balearic Island system model is developed within the EMT environment using the data conversion process mentioned above.

**Inclusion of Vendor-Specific Models and System Protection Models:** The EMT model is further enhanced by integrating vendor-specific models, such as the LCC-HVDC and VSC-HVDC links, solar PV, Distributed Energy Resources (DER), BESS, and Synchronous Condensers. Additionally, comprehensive system protection models, including generator protection and load shedding schemes, are incorporated to enhance the simulation's accuracy to real-world behaviour. This enhanced model allows for the detailed analysis of EMT and dynamic behaviors that may not be captured in the RMS environment.

**Initialization and Integration of Vendor-Specific Models for System Simulation:** The initialization of the power system for steady-state simulation employs a switch-over technique to integrate all the models with the network. The approach applies to both generic and vendor-specific models, as depicted in Figure 3, the process begins by separately initializing the network and the vendor-specific models using Thevenin voltage sources to match pre-defined load flow conditions. Subsequently, a transition occurs where the vendor-specific models are switched into the operational network, replacing the Thevenin sources. This is accomplished using three switches, with integration $T_{\text{integration}}$ marking the precise moment of integration. It is noted that reaching a new steady state post-integration varies for different models, necessitating carefully staggered integration timings to avoid excessive transients and potential converter trips. Once all models are successfully incorporated and the system attains a new steady state, dynamic simulations can commence. These are based on a saved snapshot of a single load flow condition, allowing for the exploration of various fault scenarios. The simulation workflow, including the integration process and dynamic studies, can be automated using Python scripts, ensuring efficiency and repeatability.

**Re-dispatch Process:** The final stage involves a re-dispatch process to adapt the network topology and operational conditions to reflect future developments. By doing so, the model maintains relevance not only as a current system representation but also as a dynamic entity capable of undergoing analysis for future operational scenarios.  This procedure can be achieved through the conversion tool and/or dispatch scripts.

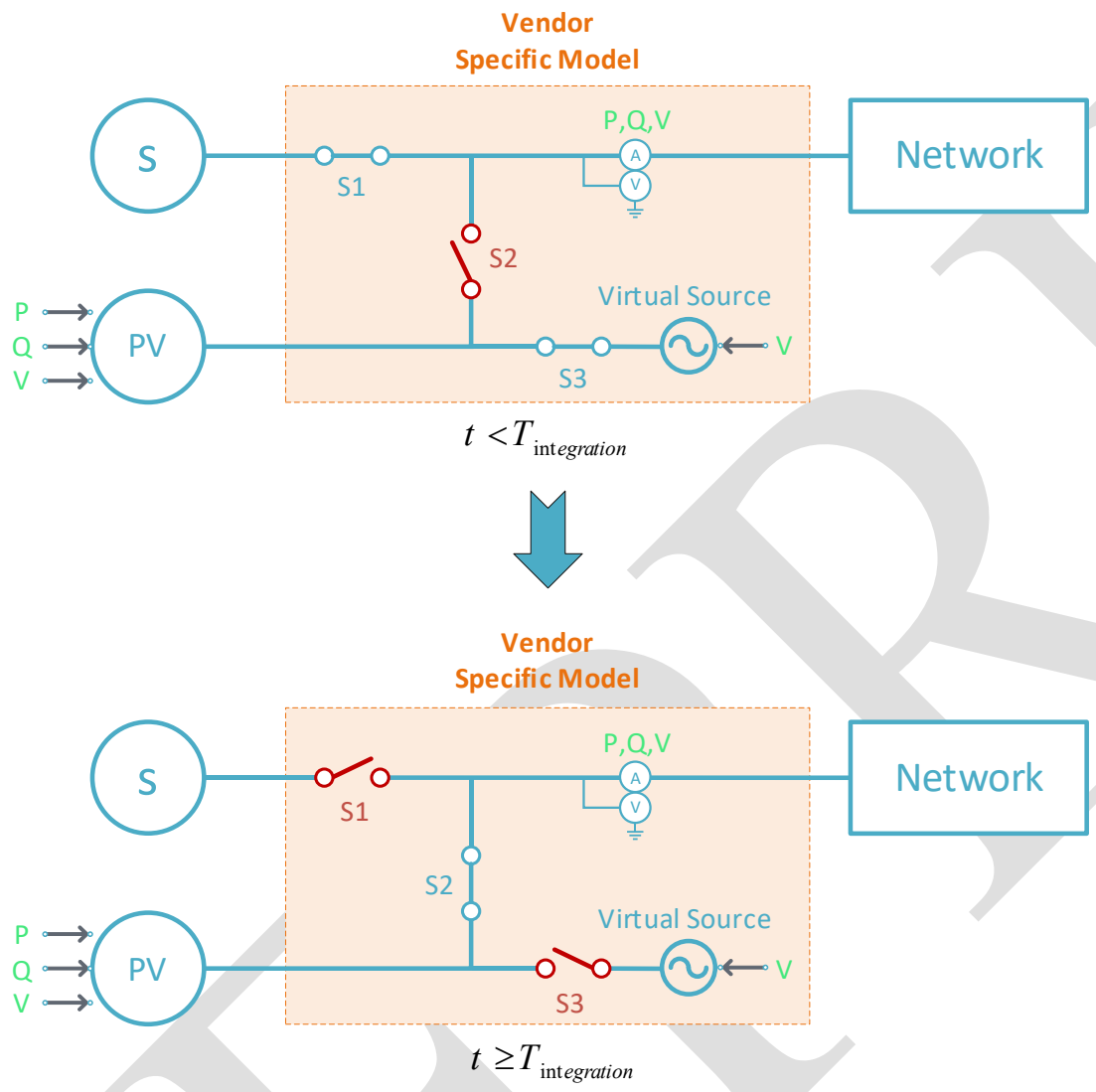

Figure 3: Schematic of the Switch-Over Initialization Process for Integrating Vendor-Specific Models into the Network

## 3. BALEARIC ISLANDS EMT MODEL WITH LARGE AMOUNTS OF INVERTER-BASED MODELS

In the previous section, the conversion process from RMS to EMT that to develop the Balearic Islands was outlined.

The workflow from the previous section was adopted to convert the Balearic Islands from RMS environment to EMT environment. This section outlines the wide-area EMT model comprehensively representing the dynamic behavior of the Balearic Islands power system. Two scenarios will be considered (SC1 of H2026 and SC2 of H2030). As an order of magnitude of the size of the EMT model of the Balearic power system, the SC1 (H2026) has 291 buses and SC2 (H2030) has 298 buses.

The EMT model segments the Balearic power system into three zones—Menorca, Mallorca, and Ibiza & Formentera—each represented as a separate project case of the EMT tool. These zones are interconnected through the Parallel Network Interface (PNI), as depicted in Figure 4.

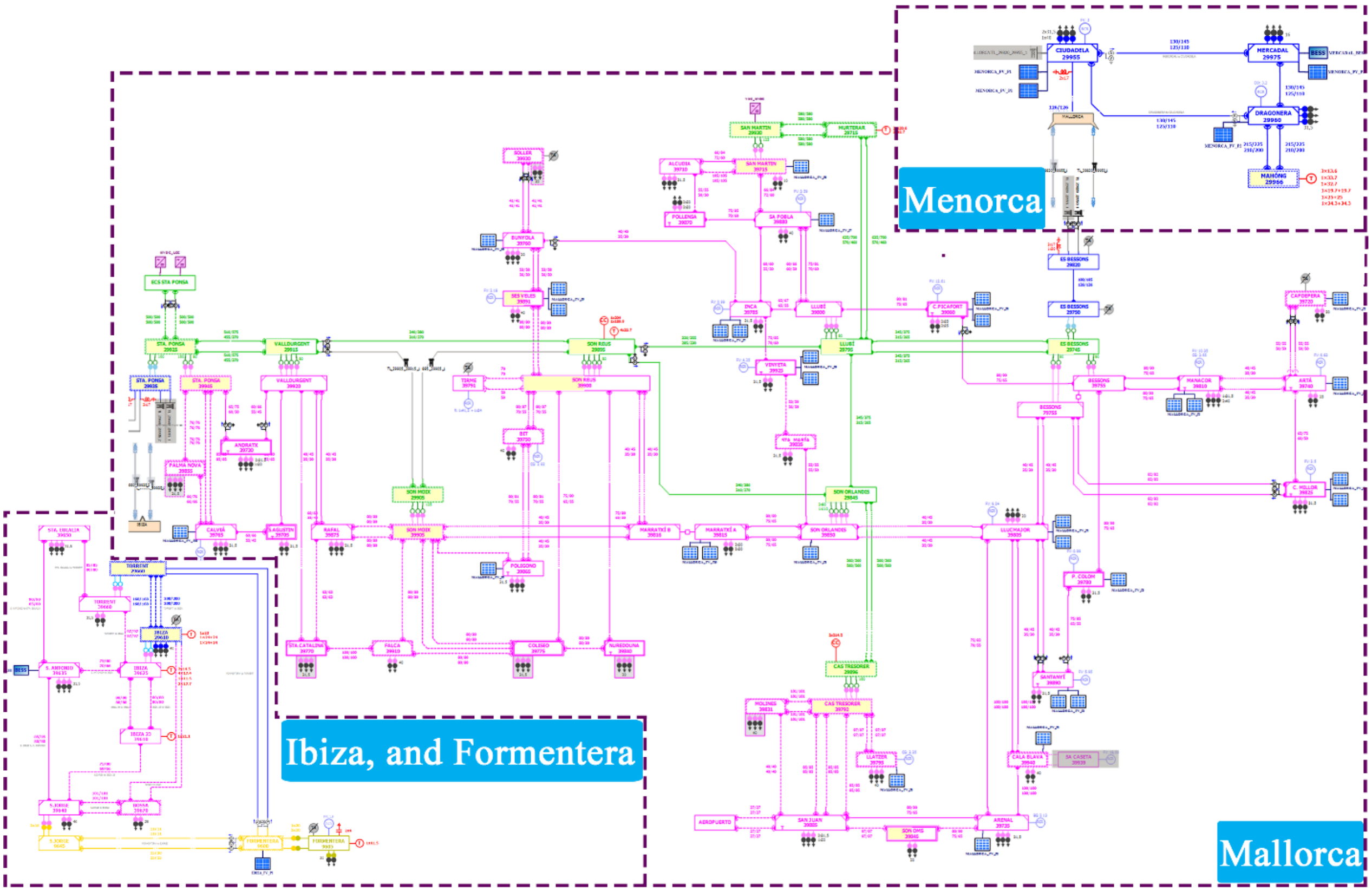

Figure 4: Illustration of the Balearic Islands Power System EMT Model Implementation.

The network model development within the EMT tool is based on a systematic connection of subnet models and vendor-specific models through Parallel Network Interface (PNI) lines. This integration method allows for an efficient simulation process, where each model is allocated to a separate page of the EMT-tool environment, leading to more effective run times during parallel simulations across multiple cores.

To represent interconnections with the mainland, the Iberian Peninsula is modelled as an equivalent 400 kV network, forming the boundary condition for HVDC links within the simulation environment. This EMT study approach, while serving current system snapshot, is also designed to be adaptable for future system studies. It includes the flexibility to expand and modify the network topology through tools like PRSIM and Python-based scripts, accommodating an increased number of vendor-specific models as dictated by planning data for future scenarios, such as the year 2030.

The computational demands are met by employing a High-Performance Computer (HPC), which enables the completion of a 10-second EMT simulation in approximately 10 minutes. This timing is subject to change depending on the hardware setup and potential optimizations in the division of network models and vendor-specific models.

Components of the Balearic Island EMT Environment include:

- **Network Representation**: A detailed model of the Balearic electrical system's network.
- **Synchronous Machine**: Custom models for synchronous machines developed and validated against the RMS environment (synchronous generators and synchronous compensators).
- **VSC-HVDC Link**: The VSC-HVDC link capable of grid-following (GFL) or grid-forming (GFM) modes at specific converter stations.
- **LCC-HVDC Link**: A model based on proprietary vendor-specific data.
- **Mainland Side (Iberian Peninsula)**: A simplified Thevenin equivalent for the external network's

boundary conditions.

- **Inverter Based Resources**: Generic models for power electronic devices, such as BESSs, and PV solar plants.
- **Protection Models**: Specific protection models developed in EMT and integrated into the system, according to grid-code requirements.
- **Special Protection Schemes (SPS)**: EMT-developed load shedding schemes integrated into the overall system.
- **Future Operating Conditions**: Initialization of models for 2026 and 2030 scenarios with steady-state points from the PSS/E power flow solution.
- **Simulation Time Step**: Typical time step ranging from 10-25 µs for vendor-specific models.

The EMT model enables system-level studies, offering valuable insights into controller interaction, transient behaviour, and system dynamics within the Balearic Islands power system. It facilitates a detailed analysis of the dynamic response and interaction between different components, ensuring grid stability, reliable operation, and seamless integration of renewable energy sources.

## 4. VSC-HVDC LINK MODEL INTERCONNECTING BALEARIC ISLANDS TO THE IBERIAN PENINSULA

A new HVDC link based on the voltage-sourced converter (VSC) technology will interconnect the Iberian Peninsula (Mainland) and the Balearic Islands (Island). The link will be a bipole system capable of transferring 400MW at a DC voltage of ±250 kV. A submarine cable link of an approximate length of 363km connects the two terminals.

A conceptual single-line diagram of the HVDC VSC system is shown in Figure 5.

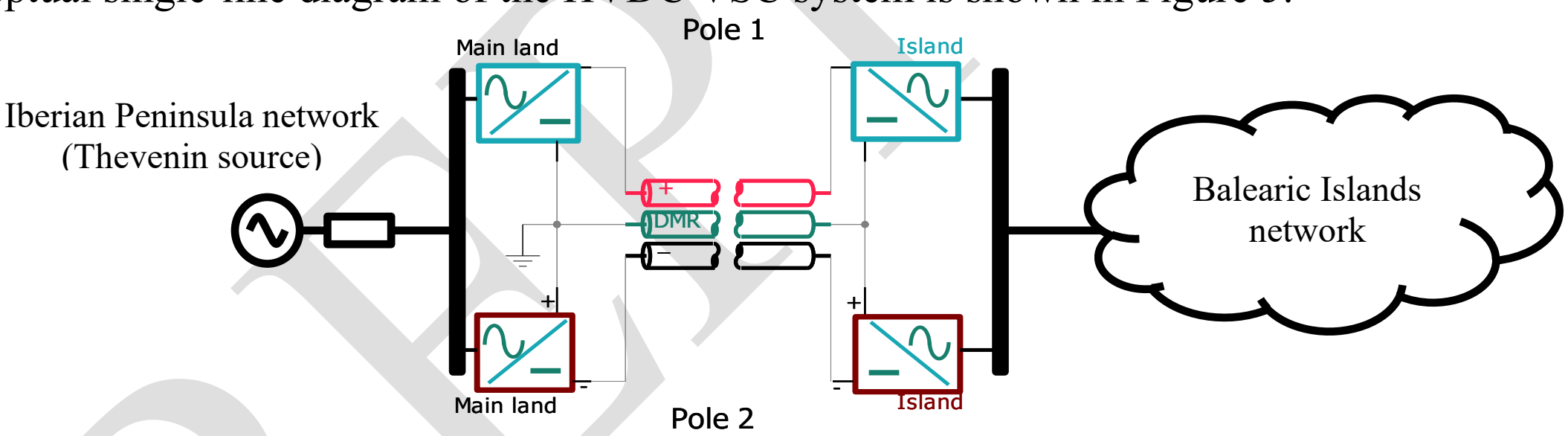

Figure 5: Single-line diagram of the HVDC VSC link between Iberian Peninsula and Balearic Islands.

A model of this VSC-HVDC system was implemented in the EMT tool. The main specifications for each pole converter are summarized in Table 1. Notice that the VSC-HVDC link has not been built yet, and some data of the model have been assumed, based on engineering judgement.

Table 1: Summary of the pole specifications

| | Each Mainland pole VSC | Each Island pole VSC |
|---|---|---|
| Topology | Half-bridge MMC | Half-bridge MMC |
| Maximum active power | 200 MW | 200 MW |
| Base power | 230 MVA | 230 MVA |
| Rated DC voltage | ±250 kV | ±250 kV |
| Grid-side voltage | 400 kV | 220 kV |
| Valve-side voltage | 160 kV | 160 kV |
| Nominal frequency | 50 Hz | 50 Hz |
| Transformer reactance (based on 230MVA) | 0.1 pu | 0.1 pu |
| Arm reactor | 50 mH | 50 mH |
| Number of submodules | 80 | 80 |
| Submodule capacitance | 7000 µF | 7000 µF |

The functions included in each of the Mainland and Island VSC terminals are explained in the following subsections.

## CONVERTERS IN THE IBERIAN PENINSULA (MAINLAND TERMINAL)

The Mainland terminal comprises two VSC poles connected to El Fadrell bus at 400 kV (Castellón, Iberian Peninsula). The converter control strategy is based on the conventional grid-following scheme described in [13] for the DC Grid Test System. The pole converters operate in grid-following mode and maintain the DC link voltage. At their AC sides, they can also operate in either reactive power control mode or in AC voltage control mode with reactive power droop. In this work, the Mainland grid is represented by a Thevenin source. The converter controllers were tuned to operate stably for short circuit ratios as low as 3.5 at the 400kV interconnection point.

## CONVERTERS IN THE BALEARIC ISLANDS (ISLAND TERMINAL)

At the island side, the VSC-HVDC link is connected to San Martín 220 kV bus (Mallorca Island, Balearic archipelago). The Island pole converters must be capable of operating in a weak system comprising several grid-following generating plants and very few synchronous generators. The Island terminal should also be able to black-start the Balearic Island grid. Therefore, grid-forming control is expected to be the preferred option for the Island terminal. In the EMT model developed in this work, the VSC-HVDC link have both options at the island terminal: (a) GFL and (b) GFM control modes and one of the two should be selected by the user. This allows to have a flexible model that could be used to analyse different situations.
The Island bipole controls the AC voltage and frequency at its AC bus (GFM control). The two pole converters operate are connected to a common AC bus. To coordinate active and reactive power sharing between the two converters and with other generators in the island grid, the following functions were included in the grid-forming controls [14]:

- Frequency droop and inertia emulation, to coordinate active power sharing.
- AC voltage control with reactive power droop, to coordinate reactive power sharing.

In case of power deficiency (or surplus) in the island network, the VSC HVDC link automatically imports power from (or exports power to) the mainland system.

A synchronization scheme controls [14] was implemented in both Island poles to allow each converter to be synchronized with the common AC bus. This is critical when the converters are to be brought into service, 1) while the island grid is already energized; or 2) in black start operation, when the two pole converters need to be paralleled.

For successful and realistic fault ride-through (FRT) performance at the Island terminal, a valve current limiting strategy was incorporated into the grid-forming controls.
To demonstrate the performance of the Island terminal, two sets of simulation results are provided in Figure 6. The following disturbance scenarios are shown in the plots:

**Test 1 – Load step changes:** A load step of 100 MW + j60 Mvar (approximately 8% of the peak demand) is connected at t = 7.5 s and disconnected at t = 10.5 s.
It can be observed that the converter successfully settles at the new operating points after each disturbance. The active and reactive power levels are evenly shared by the two poles in steady state.

**Test 2 – Fault ride-through:** A double-phase to ground fault is applied to the main 220 kV bus (AC connection point of the VSC). The fault is cleared after 150 ms. Prior to the fault, the converter supplies 200 MW + j120 Mvar to the island system.
It can be observed that the converter successfully rides through the fault. The valve currents are limited to 1.5 kA during the fault. After fault clearance, both pole converters resume supplying power to the network at the pre-fault levels.

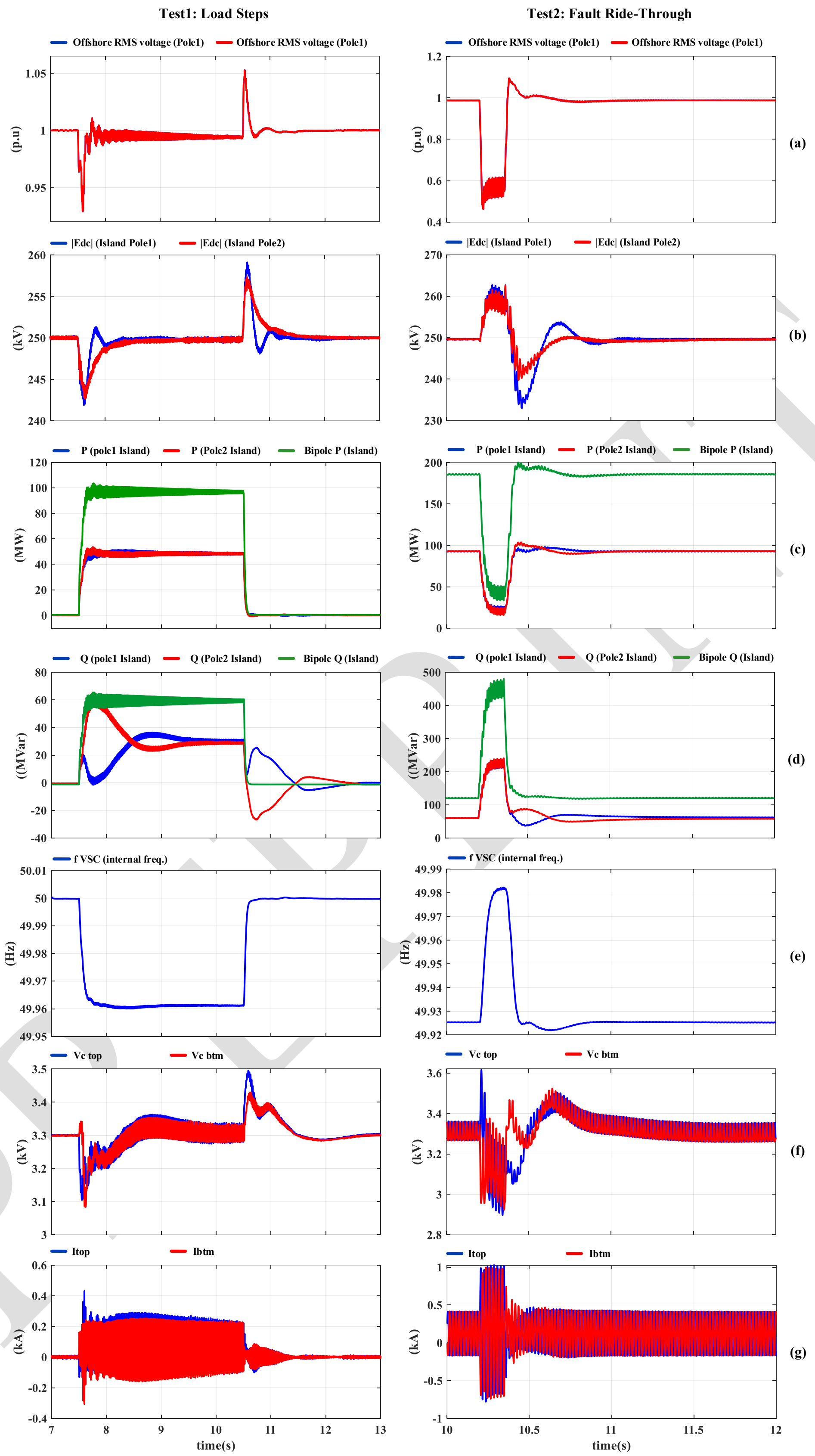

Figure 6: Dynamic performance of the Island HVDC converters. a) RMS voltage at the common bus [pu], b) Pole DC voltage magnitudes [kV], c) Active power for pole 1 (blue), pole 2 (green) and bipole (red) [MW], d) Reactive power for pole 1, pole 2 (green) and bipole (red) [Mvar], e) Frequency [Hz], f) Average submodule voltages (all valves) [kV], g) Valve currents (all valves) [kA].

## 5. STEADY-STATE AND DYNAMIC VALIDATION ACROSS RMS AND EMT ENVIRONMENTS

This section examines the dynamic response of the Balearic power system as modeled in both RMS and EMT environments under system faults. The steady-state comparison of load flows and short circuit levels (fault levels) between these environments exhibit an excellent match, affirming accurate data conversion of the electrical networks.

### VALIDATION OF INDIVIDUAL GENERATORS

The process of validating the synchronous machines and their controls usually is based on a Single Machine Infinite Bus (SMIB) setup, as shown in Figure 7. Results obtained in RMS and EMT environments are compared. The generator's conversion utilizes a file with the dynamic data of the RMS model. However, due to the non-standard nature of some models (some PSS, GOV, and any special protection features) need to be developed block-by-block in the EMT tool. The schematic includes a generator, transformer, and a Thevenin equivalent grid at the point of connection (PoC). The grid impedance is represented by an RL circuit, factoring in a short circuit ratio of 5 and an X/R ratio of 10. A three-phase solid fault is applied at the at connection point (bus 101), occurring at 2 seconds and lasting 0.1 seconds. Results depicted in Figure 8 show good agreement between RMS and EMT models. This procedure validates all synchronous machine models before integrating them into the comprehensive Balearic system model.

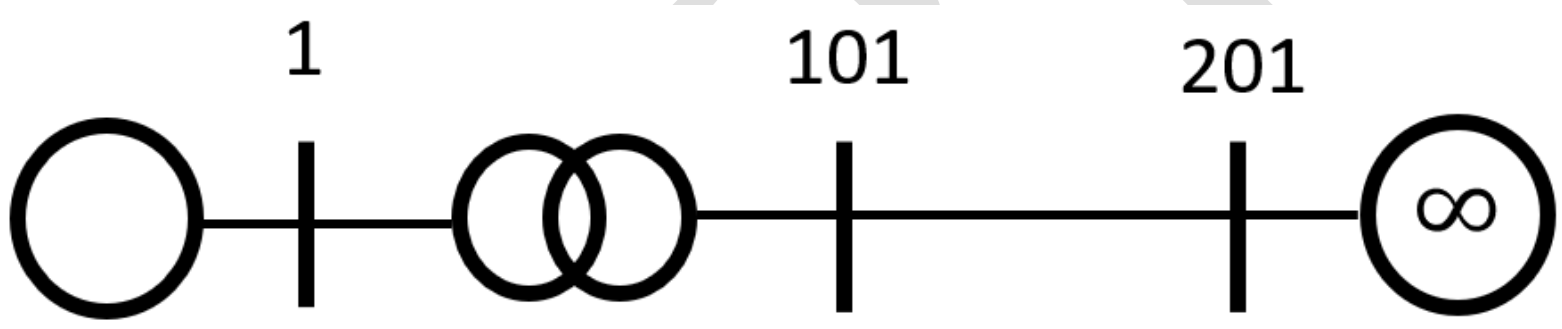

Figure 7: Test scenario for generator validation test.

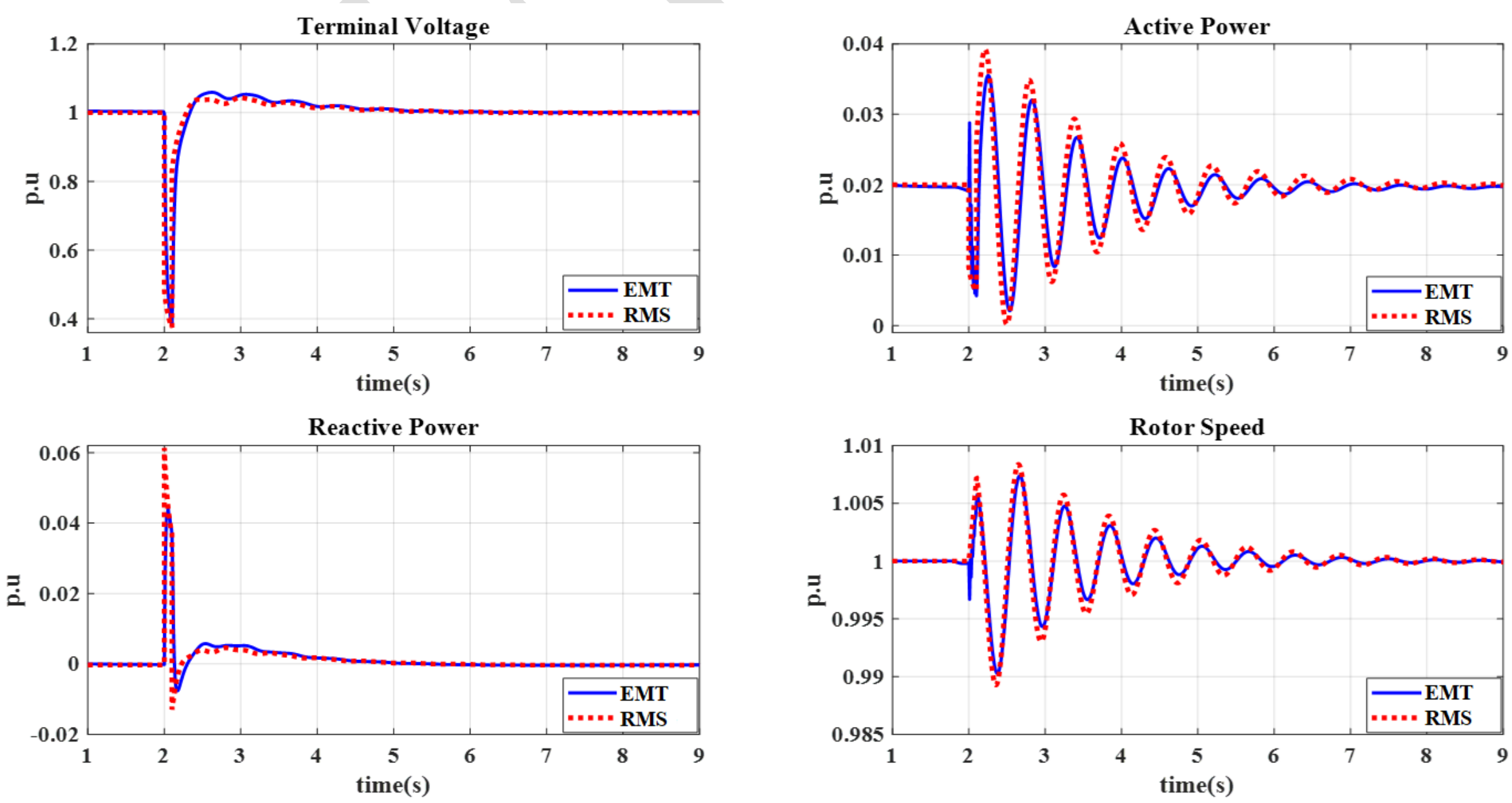

Figure 8: Test scenario for generator validation test. Validation results. Red traces: RMS, Blue trace: EMT.

## VALIDATION OF THE BALEARIC ISLANDS EMT MODEL

After the inclusion of all individually validated synchronous machine in the Balearic Islands EMT model, the EMT model is further enhanced by inclusion of all details as mentioned in Section 3. The results obtained with the EMT model of the Balearic power system are compared now with the RMS model. Naturally, the RMS model contains lower level of detail than the EMT model. As described in Section 2, the RMS model has been used as the starting point. Hence, it is important to validate the EMT model against the RMS, independently that the former contains higher level of detail than the latter. In order to make a fair comparison, the VSC-HVDC system is operating in GFL control mode at Mallorca converter station (island terminal). This is due to the fact that the generic model used for the VSC-HVDC link in PSS/E only has GFL control.

A key distinction lies in the dynamic models used for the VSC-HVDC link: in RMS it uses a simplified generic model, whereas EMT incorporates a detailed VSC-HVDC model. In the case of the LCC-HVDC links, both environments employ vendor-specific models. The main goal here is to ascertain whether both environments yield correct general trends in dynamic response whilst understanding the level of modelling fidelity between the two environments for some components are not identical. Non-synchronous generators are represented by generic models in both tools.

A 3-phase fault at a 220kV bus in Mallorca is applied that is cleared after 100 ms. Power injections of the two HVDC links (LCC and VSC) and voltage measurements at their terminals are shown in in Figure 9. Differences in the dynamic response obtained with the two models are due to (a) intrinsic differences between some models used in both tools and (b) intrinsic differences due to the level of detail captured by each type of modelling approach (RMS versus EMT). Despite variations in response details, the overall system behavior is consistent across both environments, especially during the fault, the temporary reduction in active power, gradual recovery, and reactive power behavior pre- and post-fault. However, closer examination of the fault and immediate post-fault periods reveals distinct differences between the tools. In RMS, voltage and power values may instantly change between simulation time-steps. In contrast, EMT requires a time period to reflect changes in phase quantities on RMS values. Notably, post-fault, the RMS environment instantly responds to the voltage controller, which has accumulated error during the low voltage fault, leading to an immediate reactive power change (as observed in the reactive power response in the VSC-HVDC link). In the EMT environment, due to the detailed representation of the converter, the phase-locked loop (PLL) of the firing controls takes time to resynchronize to the new system phase angle, causing a delay in the output of reactive and real power.

These differences highlight the variations typically observed between RMS and EMT tools, especially in control representation detail. As system strength diminishes and control complexity increases with a higher number of inverters, the EMT environment becomes increasingly crucial for detailed stability studies, such as control interaction, sub-synchronous oscillations, harmonic stability, etc…

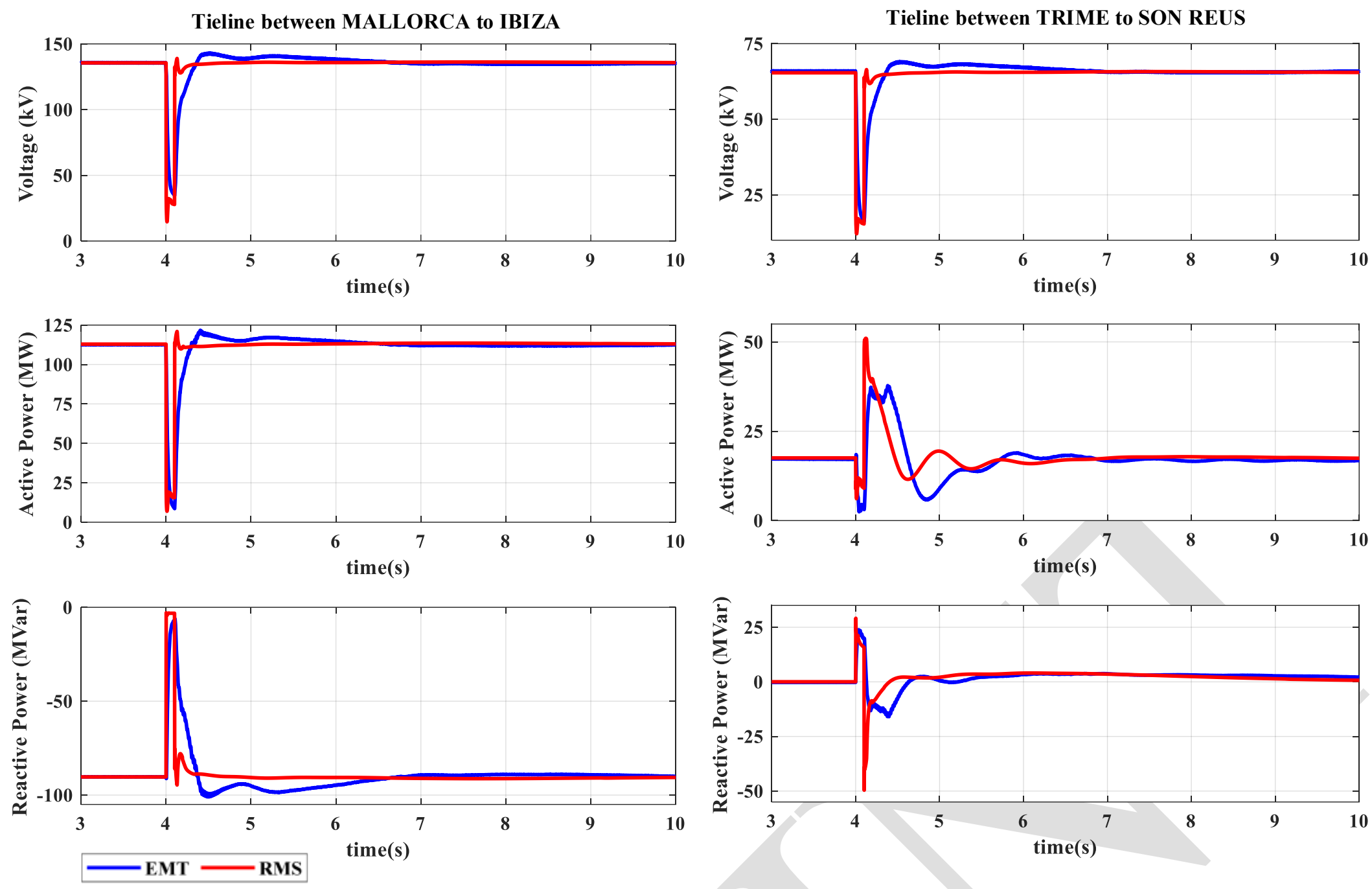

Figure 9: Voltage and Power measurements from the VSC-HVDC link (left plots) and LCC-HVDC link (right plots) in the Balearic Island. 3-phase fault at a bus that is self-cleared after 100ms in Mallorca. Red traces: RMS, Blue trace: EMT.

## 6. CONCLUSIONS

The transition to power systems with large amounts of power-electronic devices makes it essential to use EMT models to assess power system stability capturing fast dynamics such as interactions of different devices and elements in the system, as illustrated in the Balearic Islands' power system. reflects a critical shift in power system dynamics analysis in the face of increasing IBR penetration.

The following conclusions have been obtained:

1. **RMS models versus EMT models in power systems with high IBR penetration**: Although RMS models are appropriate for transient stability and frequency stability studies, they are not able to capture stability issues related to faster phenomena, such as interactions of IBR and/or the grid. In these cases, EMT models are more appropriate. The Balearic Islands' experience underlines need of EMT models to capture accurately the fast dynamic behavior of systems with significant IBR penetration.
2. **Adoption and Application of EMT Models**: The Balearic Islands' network, with its complex blend of dynamic models and system size, serves as a case study for the successful implementation and utility of large-scale EMT models. Such models are crucial for applications like generator connection studies, particularly in regions experiencing rapid IBR integration, or other type of interaction studies.
3. **Challenges in Long-Term Planning and Model Development**: Reflecting on the Balearic Islands' scenario, the limitations in long-term planning due to uncertainties in future technology and network changes become apparent. The systematic development of EMT models, necessitating clear and upfront requirements, is vital for effective modeling.
4. **Generic vs. Vendor-Specific Models**: The case of the Balearic Islands highlights the practicality of using generic models for long-term planning studies when vendor-specific models are not available, given the rapid evolution of technology and network configurations.
5. **Resource and Timeline Considerations**: The development and implementation of large-scale EMT models in the Balearic Islands underscore the significant engineering and computing resources required. It is critical to use high performance computing to ensure high efficiency.